\documentclass[12pt]{article}
\usepackage{graphicx}  
\begin{document}
\baselineskip=18pt
\leftmargin=-1.0in
\hoffset=-0.7in
\topmargin=-0.1in
\setlength{\textheight}{9.5in}
\title{Analytic formulas for frequency and size dependence of absorption 
and scattering efficiencies of astronomical polycyclic aromatic hydrocarbons}
\author{Ashim K. Roy$^{1}$\thanks{E-mail: ashim@isical.ac.in (AKR); 
sharma@bose.res.in (SKS); rag@iucaa.ernet.in (RG); pritesh@iucaa.ernet.in (PR)} ~Subodh 
 K. Sharma$^{2}$~ Ranjan Gupta$^3$ and Pritesh Ranadive$^3$\\\\
$^{1}$Indian Statistical Institute, 203, BT Road, Kolkata 700108, India\\
$^{2}$ S N Bose National Centre for Basic Sciences, Kolkata 700098, India\\
$^{3}$ Inter University Centre for Astronomy and Astrophysics, \\Ganeshkhind, Pune 411007,  
India}

\maketitle
\label{firstpage}
\newpage
\begin{abstract}
In a series of two recent papers, the frequency and size distribution dependence 
of extinction spectra for astronomical silicate and graphite grains was 
analyzed by us in the context of MRN type interstellar dust models. These 
grains were taken to be homogeneous spheres following the power law 
$(a^{-3.5})$ size distribution which is very much in use. The analytic formulas 
were obtained for the graphite and silicate grains in wavelength range 
1000\AA~-~22,500\AA~ and their utility 
was demonstrated. In this paper of the series, we present analytic formulas 
for the scattering and absorption spectrum of another important constituent   
of  interstellar dust models, namely, the polycyclic aromatic hydrocarbons 
(PAHs). Relative contribution of the PAHs to extinction {\it vis~a~vis} 
carbonaceous classical grains has been examined. \\\\

\noindent
Keywords: Interstellar grains, Extinction spectrum, PAH, Size distribution dependence.

\end{abstract}

\newpage
\noindent

\section{Introduction}
In a recent series of two papers, we presented analytic formulas for the 
frequency and size distribution dependence of the extinction spectrum of 
astronomical silicate and graphite grains in the wavelength range 1000 \AA~-
22,500~\AA~ [1,2] (hereafter referred to as RSG-1 and RSG-2). These 
formulas were worked out keeping in mind the two component interstellar dust 
models of Mathis, 
Rumple and Nordsieck (MRN) [3] type, wherein one assumes the dust grains to be 
a collection of bare 
silicate and graphite homogeneous spheres, each of the two components obeying a power law size distribution 
with exponent $-3.5$, having a minimum radius $(a_0)$ and a maximum radius 
($a_m$). In our analysis, the admitted ranges of $a_0$ and $a_m$ were 
$0.002~\mu m\leq a_0\leq 0.005~\mu m$ and $0.15~\mu m\leq a_m\leq 0.25~\mu m$ 
for graphite grains and $0.004~\mu m\leq a_0\leq 0.006~\mu m$ and 
$0.2~\mu m\leq a_m\leq 0.4~\mu m$ for silicate grains. These ranges are thought to be quite appropriate and allow one to have the kind of flexibility needed 
in fixing the values of $a_0$ and $a_m$. A comparison of 
extinction predicted by these formulas {\it vis-a-vis} exact Mie computations 
showed that these formulas give results which are in close proximity to exact 
computations. Hence, these formulas have the potential to be used efficiently 
for assessment of the extinction contributions of the silicate and graphite 
components in investigations and building of MRN type models without resorting 
to large scale numerical computations. Moreover, by selecting proper ranges of 
frequency, the extinction behavior can be effectively approximated by yet 
simpler forms (e,g., linear, quadratic, etc.). That is, further simplification 
can be made in an effort to set appropriate particle size ranges to start 
with. It may be mentioned here that analytic formulas fitting the extinction 
spectra of 
several stars have been obtained in the past by Cardelli {\it et al.} [4] and 
Fitzpatrick and Massa [5]. However, these parametrizations are mostly based on 
mathematical schemes. In contrast, the formulas provided in RSG-1 and RSG-2, 
give some degree of physical insight into the problem and also address the 
problem in terms of silicate and graphite grains separately with a provision 
of allowable variations in $a_0$ as well as in $a_m$ - which are known to be 
quite adequate.\\\\ 
In the MRN type models it is generally assumed that carbonaceous grains have 
$a_0\approx 0.005~ \mu m$ and their optical 
properties are more like graphite 
grains. However, in more recent models it has been recognized that the 
carbonaceous component also has a sub-component of ``very small grains" whose 
optical properties are different from that of graphite. These grains are the  
so-called polycyclic aromatic hydrocarbons that are referred to as ``astronomical"
PAHs in the literature.\
\subsection{Why PAH's}
In 1956 Platt[6] proposed that very small grain or large molecules of radii around 
10\AA~ to be present in the Interstellar Dust. Donn(1968)[7] proposed that polycyclic aromatic hydrocarbon 
like 'Platt Particles' may produce UV extinction in the interstellar spectra.
Importance and ubiquity of polycyclic aromatic hydrocarbons is quite evident in literature addressing the
interstellar extinction. The inclusion of PAH's is based on the three decades of research by various groups.
The IR emission features  are very well attributed to PAH's by now and
also allow to place constraints on size distribution of very small dust components.
Ubiquity of PAH's in many galaxies has been observationally verified by ISO and Spitzer, thus making Polycyclic 
Aromatic Hydrocarbons (PAH's) a very important constituents of Interstellar Dust(Draine and Li 2007,[8]).
Draine and cowrokers (Li and Draine 2001b[9], 2002a[10]; Weingartener and Draine 2001a[11]) proposed the Silicate-
Graphite-PAH's model, of which PAH are the small-size end of the carbonaceous grains, is in excellent agreement 
with observations.For more details please refer to the paper by Li and Greenberg[12]. On these lines, in this paper, we aim at extending the analyses of RSG 1 and RSG 2 
to include the extinction spectra analysis of the  
PAHs in the size range 5 \AA~$\leq a\leq $ 50 \AA~ and spectral domain  
1000 \AA~ to 22,500 \AA~. The closed form analytic expressions for 
extinction by such grains should enable one to make more precise and elaborate 
model investigations and hence correct model building. The PAHs like grains 
can be further distinguished as being either in neutral or ionic state. 
Analytic formulas to evaluate and analyze the extinction spectrum have been 
obtained for both types of PAHs in this work. As the PAHs are very small in  
size, the scattering is expected to be unimportant [13]. Thus, the  extinction 
can be effectively computed from the absorption only. However, for 
completeness, formulas have been obtained for absorption as well as scattering efficiencies.\\\\  
With this paper of the series, we aim to complete our objective of 
constructing an analytic framework by the use of which a direct approach to 
analysis of interstellar dust extinction spectra available for various 
galaxies {\it viz.~} MW, LMC, SMC~ etc. can be done very expediently. Whereas, 
it would be our endeavor to carry out the observed interstellar dust 
extinction spectra data analysis in context of MW, etc., in a forthcoming paper,
nevertheless, a comparison of the extinction produced by equal volumes of 
PAHs and the two varieties of graphite (perpendicular and parallel) separately 
is examined here to appreciate the important role of PAHs in shaping the 
extinction spectra of the dust containing them. The mass density of PAHs and 
graphite is taken to be the same ($\approx 2.24~ gm/cm^3$) here, as per the 
general convention so 
that equal volume also implies equal mass. \\\\
The absorption and scattering efficiency data, covering a wide radius 
range~ $0.000355\mu m\leq a\leq 0.01~\mu m$~ and the wavelength 
range ~$10~\AA~\leq\lambda\leq ~10^7~\AA$, for the ``astronomical" PAHs in excellent 
tabular form can be found on the website of Draine [14]. The details in this regard are 
available in the appendix part of [9]. These results, to be 
parametrized in this paper, are referred to as the exact results. \\\\
This paper has been organized as follows. Section 2 describes all the relevant 
functional forms of scattering and absorption efficiencies obtained for PAHs 
in the wavelength range $1000 \AA~\leq\lambda\leq ~22,500 \AA$  for size range 
5 \AA~$\leq a\leq $ 50 \AA. The accuracy of all the formulas obtained has been 
demonstrated by numerically computing the absorption and scattering 
efficiencies from the formulas and contrasting them with the corresponding 
exact results of Draine [14]. It may be mentioned that the parametrization of 
these so-called exact results in various wavelength domains have also been 
achieved by Li and Draine [9]. A typical numerical comparison of their 
formulas with exact absorption efficiencies has also been presented here in 
the UV and 
FUV regions. In section 3, the extinction spectra of various constituents of 
the carbonaceous component of the dust has been discussed and their relative 
contributions to 
the extinction spectrum have been analyzed. The convenience introduced by analytic 
formalism  becomes very apparent in this part of the analysis. The dust model 
has been described contextually in this section. Finally, we conclude by 
summarizing and discussing the salient results of this paper in section 4. In 
this work, grains are assumed to be spheres and, therefore,  this work ignores 
issues related to polarization.
\section{ Parametrization of PAHs absorption and scattering efficiencies}
The computed values of scattering and absorption cross sections of PAHs  in 
the size range 3.5 \AA~$\leq a\leq $ 100 \AA~ have been tabulated by 
Draine [14] for neutral as well as for ionized PAHs in the wavelength 
range 10\AA~$\leq\lambda\leq ~ 10^7$\AA~. We have attempted to obtain 
simple parametrization for corresponding absorption and scattering 
efficiencies for PAH sizes 5 \AA~ $\leq a\leq$  50\AA~ for  wavelength range 
1000 \AA~$\leq\lambda\leq$  22,500 \AA. The PAHs in this size range have been  
taken to be the major contributors to the absorption (extinction) due to dust 
in Li and Draine [9]. Thus, it was decided to restrict the present 
investigation to the PAH size range 5\AA~$\leq a\leq 50$ \AA. The  admitted 
wavelength range is the same as 
it was in RSG-1 and RSG-2. \\\\
For absorption efficiencies, the total wavelength range has been divided into  
five regions, corresponding to far-ultraviolet 
(1000 \AA~$\leq\lambda$ 1800 \AA), ultraviolet 
(1800 \AA~$\leq\lambda\leq$ 4000 \AA), visible 
(4000 \AA~$\leq\lambda\leq$ 8000 \AA), infrared-I 
(8000 \AA~$\leq\lambda\leq$ 12,500 \AA) and infrared-II 
(12,500 \AA~$\leq\lambda\leq$ 22,500 \AA). 
The corresponding formulas have been accomplished for the neutral as well as 
the ionic PAHs.  
Barring region 4 (infrared-I), in each of the other regions, it was possible 
to arrive at a single formula covering   
the entire size range 5 \AA~$\leq a\leq$ 50 \AA. In the region 4, separate 
formulas were required for the ranges 5 \AA~$\leq a\leq$ 10 \AA~ and 10 \AA~$\leq a\leq $ 50 
\AA. For the ionic PAHs however, separate formulas are required in regions 3, 4 and
 5. In regions 1 (FUV) and 2 (UV), the corresponding absorption efficiency formulas for 
the neutral and the ionic particles are identical.
 For scattering efficiencies, we have divided the wavelength range of our 
interest in three spectral regions. Theses are far-ultraviolet 
(1000 \AA~$\leq\lambda\leq$ 1800 \AA), ultraviolet 
(1800 \AA~$\leq\lambda\leq$ 4000 \AA) and visible- 
infrared (4000 \AA~$\leq\lambda\leq$ 22,500 \AA). \\\\
A parametrization of absorption cross-section of PAHs has also been obtained 
by Li and Draine [9] starting from the far-ultraviolet to the far-infrared in 
the size domain 3.5\AA $\leq a\leq$ 100\AA. Their parametrization is 
characterized by a set of Drude profiles. Whereas for $\lambda\leq$3300\AA~ 
the formulas require at most one Drude profile, as many as 12 Drude profiles 
have been used to construct the formula for $\lambda\geq $ 3300 \AA.\\\\
In this work, numerical comparisons have been generally presented for 
the values of radius $a=10,~20,~35$ and $50$ \AA. But, it may be mentioned 
that, in addition, we have also verified all the results for the intermediate 
values of $a=5,~8$, $15$
 and $45$ \AA~ to ensure that all observations and statements made based on 
them are really speaking valid over the size range 5\AA~$\leq a \leq$50\AA.
\subsection{Neutral PAHs}
In all the formulas to follow, the radius ($a$) and the wavelength ($\lambda$) 
are in the units of $10^{-5}~ cm$ unless stated otherwise. The choice of units
is made to confirm with our earlier work (RSG-1 and RSG-2) wherein, the same 
units are used in the formulas for extinction spectra of the silicate and the 
graphite particles. The advantage of using these units is to have the numerics 
associated with $\nu~(=1/\lambda)$ and $a$ as coefficients or in radicals 
appearing in the various formulas are neither too large, nor too small.
\subsubsection{Absorption efficiency}
For  absorption efficiency, we arrive at 
the following formulas:\\\\
{\bf 1000\AA~$\leq\lambda\leq$ 1800\AA~ (FUV) }
$$Q_{abs}(1)= x\Biggl[\frac{0.93265}{\nu} +\frac{19.608}{\nu}(\nu-0.643)^2-
\frac{22.2817}{\nu}(\nu-0.7023)^3  +$$
$$\frac{6.0}{\nu}(\nu-0.7023)(1.0-\nu)(\nu-0.55)\Biggr],~~~(3.5 \mbox{\AA}\leq a \leq 50\mbox{\AA}) \eqno (1)$$
where $x=2\pi a\nu$. Note that the absorption efficiency is 
proportional to $x$. This is as it should be.  For particles small compared to 
the wavelength of the incident radiation ($x\ll 1$), the general expression 
for the absorption efficiency, as a power series in $x$, is known to be of 
the form (see, for example, Pendorf [15]):
$$Q_{abs}(x)\approx x (A+Bx^2 +Cx^3+.....). \eqno(2)$$
where $A,~B,~C$ etc. are functions of the refractive index of the scatterer 
relative to the refractive index of the surrounding medium. 
The minimum and maximum values of $x$ in the present study are $0.000977$ 
(corresponding to $a=3.5$ \AA~; ~ $\lambda=$ 22,500 \AA~) and $0.314$ 
(corresponding to $a=50$ \AA~; ~ $\lambda=$ 1000 \AA~) respectively. 
Therefore, in obtaining the formula 
(1) the first term of the general expression (2) is found to be sufficient in 
parametrization of the exact absorption efficiency successfully. Higher order 
terms in $x$ give negligible contribution. \\\\
Li and Draine [9] have given two formulas for the absorption efficiency to 
cover this spectral region which are as follows. For        
$ 1000\AA~\leq\lambda\leq 1300\AA $,
$$ Q_{abs}=\frac{a}{100.0(1.286)^3\pi}\Biggl[66.302-24.367\nu+2.950\nu^2-0.1075\nu^3\Biggr], \eqno (3)$$
while, for $ 1300\AA~\leq\lambda\leq 1700\AA $,
$$ Q_{abs}=\frac{a}{100.0(1.286)^3\pi}\Biggl[S_2(\lambda)+1.8687+0.1905
\nu +0.4175(\nu -5.9)^2 + 0.04370(\nu-5.9)^3\Biggr], \eqno(4)$$
where $S_2$ is the particular case ($j=2$) of the general expression for 
Drude's spectral absorption profile:
$$S_j(\lambda)\equiv\frac{2}{\pi}\frac{\gamma_j\lambda_j\sigma_{int,j}}{(\lambda/
\lambda_j -\lambda_j/\lambda)^2+\gamma_j^2  }.
 \eqno(5)$$
Values of  $\gamma_j$, $\lambda_j$ and $\sigma_{int,j}$ are given in [9]. The 
expressions obtained therein are for absorption cross section 
per $C$ atom. The corresponding expression for the absorption efficiency is 
given in the appendix. \\\\\
{\bf 1800\AA~$\leq\lambda\leq$ 5000\AA~ (UV)}\\
In this wavelength region too, the parametrization of the absorption 
efficiency could be achieved in a form such that it is proportional to the size 
parameter $x$. 
$$Q_{abs}(2)=x\Biggl[\frac{1.19}{\nu}-3.99\Bigl|1.0-\frac{0.462}{\nu}\Bigr|+
\frac{1.0}{0.74\nu+83\nu(1.0-\frac{0.46}{\nu})^2}\Biggr], ~~~(3.5\mbox{\AA}\leq a\leq 50\mbox{\AA})\eqno (6)$$
The formulas obtained in [9] for this region are:
$$ Q_{abs}=\frac{a}{100.0(1.286)^3\pi}\Biggl[S_2(\lambda)+1.8687+0.1905\nu
\Biggr], \eqno(7) $$ 
for  1700\AA~$\leq\lambda\leq$ 3030\AA , and 
$$ Q_{abs}=\frac{a}{100.0(1.286)^3\pi}\Biggl[34.58\times 10^{-3.431/\nu}~\mbox{cutoff}(
\lambda,\lambda_c) +\sum_{j=3}^{14}S_j(\lambda)\Biggr], \eqno(8)$$
for  $\lambda\geq$ 3030\AA~ , where $S_j$ is given by equation (5). It may be 
noted that cutoff$(\lambda,\lambda_c)$ is a function of $a$  
as well as $\lambda$ because $\lambda_c$ is a function of the radius $a$ of the
PAH particle [9]. \\\\
{\bf 4000\AA~$\leq\lambda\leq$ 8000\AA~ (Visible)}
$$Q_{abs}(3)=x\Biggl[\bigl(28.176\nu^2-0.091\nu-0.311\bigr)-(a-0.048)^2 \Bigl(4600\nu^2 -2052.66\nu + 226.26\Bigl)\Biggl] -$$
 $$ \frac{(0.041/100a)^{5/2}}{1+(2175a)^3\Bigl[0.1228+\frac{0.01312}{(100a)^2}-\nu\Bigr]^2} - \Bigl(1.3-\frac{1}{4\nu}\Bigr)\Biggr(\frac{0.00355}{10a}\Biggr)^3 , ~~~(5\mbox{\AA}\leq a\leq 50\mbox{\AA})  \eqno (9) $$
\\
Figure 1(a) shows a comparison of absorption efficiencies computed on the basis of
our formulas with the exact results [14] for four PAHs sizes. The 
wavelength range is  1000\AA~to~8000\AA. 

\begin{figure}[!h]
\centerline{\includegraphics[width=5.0in, height=10cm]{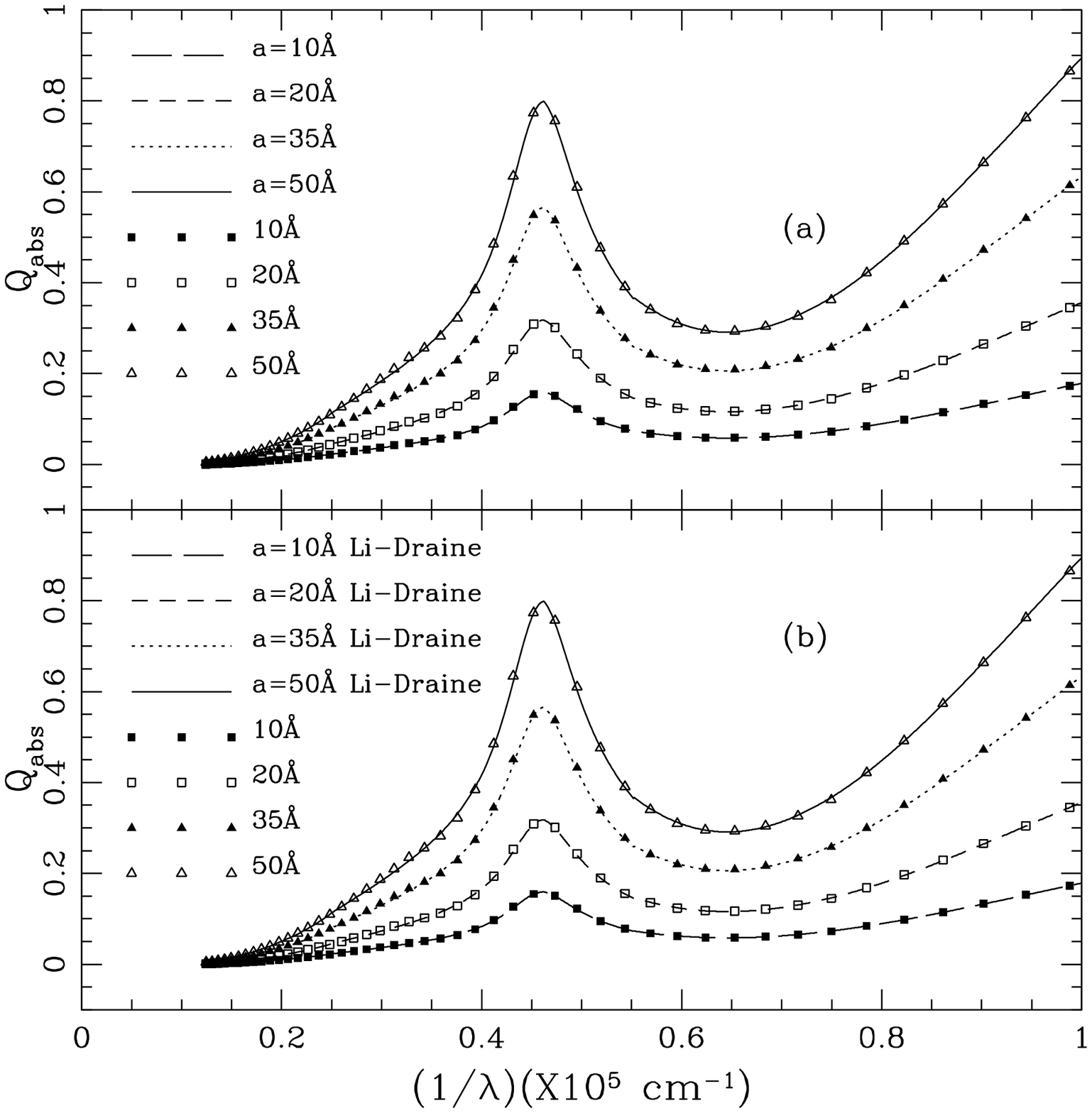}}
\begin{footnotesize}Fig1 (a): Comparison of predictions (lines) of absorption efficiencies from our equations with exact results
 (Draine website[8]) points\\
     (b): Comparison of predictions (lines) of absorption efficiencies form Li and Draine 2001 equations 
   with exact results
 (Draine website[8]) points\\
     Frequency range $0.125\leq\nu\leq 1.0$, wavelength range 1000\AA~$\leq\lambda\leq$~8000\AA (i.e. FUV+UV+Visible region) 
   \end{footnotesize}
\end{figure} 

It can be seen that the agreement 
between the predictions of (1) and (6) and the exact results is excellent. To contrast results from 
formulas (1) and (6) with those from the parametrization by Li and Draine [9], 
a typical comparison of absorption efficiencies computed from  
formulas (3), (4), (7) and (8) with exact results is shown in  Figure 1(b). 
Clearly, the accuracy of Li and Draine parametrizations and our formulas is 
almost identical. \\

{\bf 8000\AA~$\leq\lambda\leq$ 12,500\AA~ (Infrared- I)}
$$Q_{abs}(4)=\frac{a\Bigl[0.01391+a(3.1296-32.3a)\Bigr]}{1.0+\Bigl[113.3+(10.43-\frac{0.27}{a})^2\Bigr](\frac{0.1245}{\nu}-1.0)^2}~+~
\frac{(0.00422/100a)^2}{1.0+(10\nu-0.8)^2}, $$
$$~~~~~~~~~~~~~~~~~~~~~~~~~~~~~~~~~~~~~~~~~~~~~~~~~(10\mbox{\AA}\leq a\leq 50\mbox{\AA}) \eqno(10a)$$

$$Q_{abs}^s(4)=\frac{0.00208a}{1.0-2.4\Bigr[0.22+a(94.0+C)\Bigr]\Bigl(1.0-\frac{0.00644}{\nu^2}\Bigr)^2}+\Biggl(\frac{0.00117}{a}\Biggr)^4(\nu-0.08)(0.1245-\nu),
$$  $$~~~~~~~~~~~~~~~~~~~~~~~~~~~~~~~~~~~~~~~~~~~~~~
~~~(5 \mbox{\AA}\leq a\leq 10\mbox{\AA}) \eqno(10b)$$
where
$$C=438(1.0-100a)(100a-0.5)^2.$$\\
and the superscript $s$ in (10b) stands for small PAH range 5\AA~$\leq a \leq$
10\AA. Figure 2(a) exhibits a comparison of variation of absorption efficiency 
computed on the basis of (10a) with exact results. For clarity of display, the 
comparison has been shown only for two values of PAH sizes, $a=20$\AA~ and 
$a=50$\AA. The agreement can be seen to be excellent. Although not shown here, 
computations have been done for (10b) also and the results have been found to 
agree reasonably well with exact results. \\
It appears that the molecular structure of PAHs become vital in this 
wavelength range, for the absorption does not seem to follow the general 
expression (6) which is obtained from the expansion of Mie formulas for 
small $x$ values. \\\\

\begin{figure}[!h]
\centerline{\includegraphics[width=5.0in, height=10.0cm]{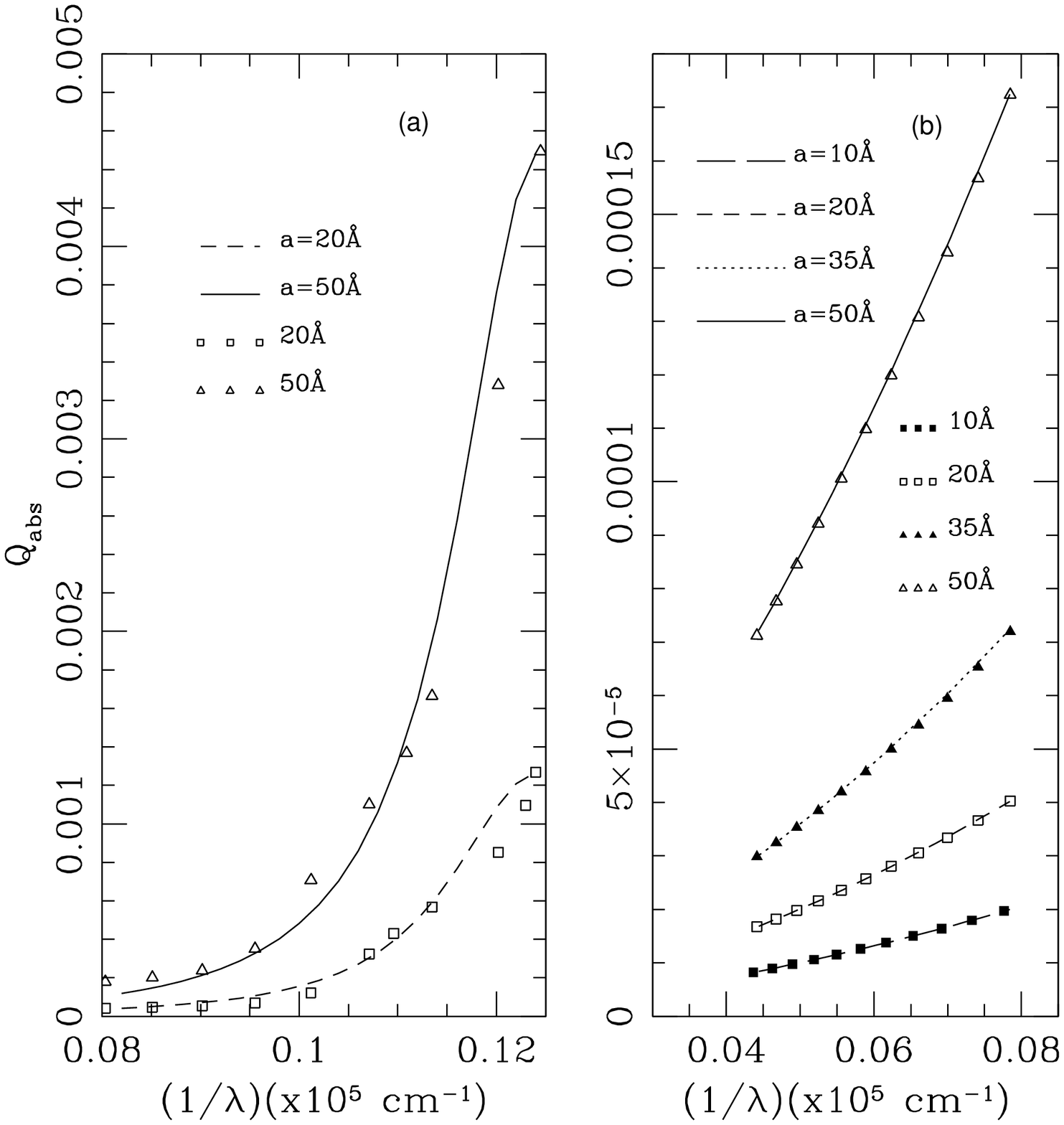}}
\begin{footnotesize}Fig 2: Comparison of predictions (lines) of absorption efficiencies of our equations with exact result
 (Draine's website [8]) points\\
 (a): Frequency range $0.08\leq\nu\leq 0.125$; wavelength range 8000\AA~$\leq\lambda\leq$ ~12,500\AA~ (IR-1)\\
 (b): Frequency range $0.044\leq\nu\leq 0.08$; wavelength range 12,500\AA~$\leq\lambda\leq$ ~22,500\AA~ (IR-2) 
\end{footnotesize}
\end{figure}

{\bf 12,500\AA~$\leq\lambda\leq$ 22,500\AA~ (Infrared- II)}
$$Q_{abs}(5)=0.01x\Biggl[\frac{\sqrt{\nu}}{1.0+(1000a-50.0)^4}+3\nu(1.0+2.2a) +0.168 -0.2a\Biggr],~~~(5\mbox{\AA}\leq a\leq 50\mbox{\AA}) \eqno(11)$$
A comparison of absorption efficiencies computed from (11) with exact results 
is depicted in Figure 2(b). Once again, the agreement of our formula with exact 
results is excellent.\\ 
Above comparisons show that our formulas for absorption efficiency give 
excellent results over the entire wavelength and size range considered here. 
Formulas obtained involve only frequency of radiation and the size of the 
scatterer and  are simple to use. In contrast, the Li and Draine formulas [9], 
for $\lambda\geq$ 3000\AA~, require contributions from as many as 12 Drude 
profiles making it somewhat cumbersome to use fluently. The formula also 
involves a function cutoff($\lambda,\lambda_c$) containing the cutoff 
wavelength $\lambda_c$ which in turn, depends on the PAH radius in a complex 
manner. 
\subsubsection{Scattering efficiencies}
For the scattering efficiencies, the formulas for the neutral or the ionic PAHs 
turn out to be  
identical. These are as follows.\\\\
{\bf 1000\AA~$\leq\lambda\leq$ 1800\AA~ (FUV)}
$$Q_{sca}(1)=x^4\Biggl[\frac{0.112}{\nu^2}+9.8\Biggl(1.0-\frac{0.620}{\nu}\Biggr)^2 + 
\frac{36.124}{\nu^2}\Biggl(1.0-\frac{0.657}{\nu}\Biggr)^4 + $$ $$x^2
\Biggl(\frac{\nu}{0.507\nu +34.0(1-\nu)^2}- \frac{x}{1.96+2.665\nu-4.2\nu^2}
\Biggr)
+\frac{0.09}{[1.0+(40\nu-38.0)^2]^2}\Biggr], 
~~~(5\mbox{\AA}\leq a\leq 50\mbox{\AA})\eqno(12)$$\\\\

{\bf 1800\AA~$\leq\lambda\leq$ 4000\AA~(UV)}
$$Q_{sca}(2)=x^4\Biggl[\frac{1.0}{\nu \bigl(0.4485+157.0\nu^2 (1-
\frac{\nu}{0.4562})^2\bigr)}+\frac{0.25}{1+(50\nu -16)^2}\Biggr]. 
~~~(5\mbox{\AA}\leq a\leq 50\mbox{\AA})\eqno(13)$$\\\\

\begin{figure}[!h]
\centerline{\includegraphics[width=5.0in, height=10.0cm]{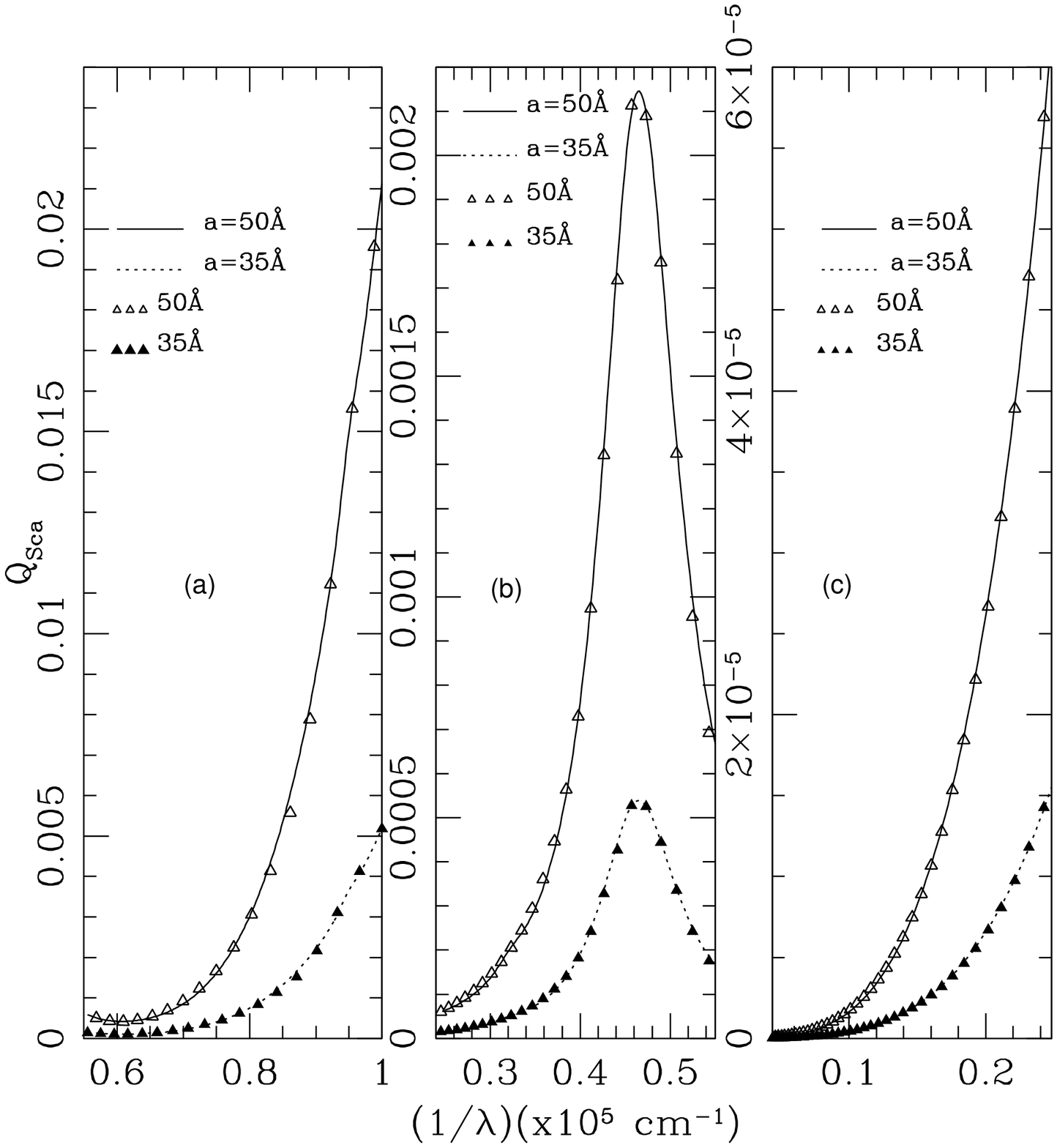}}
\begin{footnotesize}Fig 3: Comparisons of predictions (lines) of scattering efficiencies from our equations with the exact results
  (Draine's website [8]) points\\
(a): Frequency range $0.55\leq\nu\leq 1.0$; wavelength range 1000\AA~$\leq\lambda\leq$ ~1800\AA~(FUV)  \\
(b): Frequency range $0.25\leq\nu\leq 0.55$; wavelength range 1800\AA~$\leq\lambda\leq$ ~4000\AA~(UV) \\
(c): Frequency range $0.044\leq\nu\leq 0.25$; wavelength range 4000\AA~$\leq\lambda\leq$ ~22,500\AA~(Visible+IR) \\
 \end{footnotesize}
\end{figure}

{\bf 4000\AA~$\leq\lambda\leq$ 22,500\AA~(Visible,~ IR-I,~ IR-II)}
$$Q_{sca}(3)= 1.66~x^4, 
~~~(5\mbox{\AA}\leq a\leq 50\mbox{\AA})\eqno(14)$$

The functional forms of the scattering efficiency in these spectral regions 
may be contrasted with the expansion of exact $Q_{sca}$ in powers of $x$ for 
small $x$ (see, for example, Pendorf [15]): 
$$Q_{sca}(x)\approx x^4(D+Ex^2 +Fx^3 +......) \eqno(15)$$
where $D,~E,~F$ etc. are effectively, functions of the refractive index. It 
may be noted that while for (13) and (14) only the first term (of the order of
 $x^4$) is sufficient to reproduce the observed results, for (12) the second 
and the third terms, of the order of $x^6$, $x^7$ are also required.  
\subsection{Ionic PAHs}
In the wavelength regions 1 and 2 (FUV, UV) the formulas remain the same as 
that for non-ionic PAHs. For the other three regions (Visible, IR-I, IR-II), 
following formulas have been obtained for the absorption efficiencies,\\\\
{\bf 4000\AA~$\leq\lambda\leq$ 8000\AA~ (Visible)}
$$Q_{iabs}(3)=x\Biggl[28.176\nu^2-0.091\nu-0.311-(a-0.048)^2(4600\nu^2-2052.66\nu+226.26)-$$ $$\frac{(0.4858/100a)^4}{1.0+20.0(100\nu-14.3)^2}\Biggr] + \frac{0.0006(0.01/a)^3}{1.0+0.20(100\nu-13.4)^4},  
~~~ (10\mbox{\AA}\leq a\leq 50\mbox{\AA}) \eqno (16a)$$\\
$$Q_{iabs}^s(3)=a\nu(178.568\nu^2-2.184)-\frac{0.0005(0.005/a)^3}{1.0+7(100\nu^2-1.96)^2,} ~~~ (5\mbox{\AA}\leq a\leq 10\mbox{\AA}) \eqno (16b)$$
{\bf 8000\AA~$\leq\lambda\leq$ 12,500\AA~ (Infrared- I)}
$$Q_{iabs}(4)=\frac{0.1a-0.0001}{1.0+\Bigl[68.0+\frac{0.41}{a}+37.55\Bigl(\frac{0.0207}{a}-1.0\Bigr)^3\Bigr]\Bigl(\frac{0.1245}{\nu}-1.0\Bigr)^2}, ~~~(10\mbox{\AA}\leq a\leq 50\mbox{\AA}) 
\eqno(17a) $$ \newline
$$Q_{iabs}^s(4)= \frac{0.21a-0.001077-0.01(100a-0.768)^3}{1.0+\Bigl[\frac{0.90}{a}+\frac{28.0}{\sqrt{a}}|1.0-\frac{0.0071}{a}|\Bigr]\Bigl(\frac{0.1245}{\nu}-1.0\Bigr)^2} + 
\frac{(3.425-25\nu)(0.005/100a)^2}{10.0+3\Bigl(10.0-\frac{0.89}{\nu}\Bigr)^2},$$ $$~~~~~~~~~~~~~~~~~~~~~~~~~~~~~~~~~~  ~~~(5\mbox{\AA}\leq a\leq 10\mbox{\AA}) \eqno(17b)$$
{\bf 12,500\AA~$\leq\lambda\leq$ 22,500\AA~ (Infrared- II)}
$$Q_{iabs}(5)=Q_{abs}(5)+ a\Bigl[(0.005)^2+0.116(\nu-0.0675)^2)\Bigr] + 
\frac{0.00352a-(0.04134-a)^3}{1.0+\Bigl[\frac{0.22}{a}\Bigl(\frac{0.89}{\nu}-10\Bigr)^4\Bigr]}, $$  $$
~~~~~~~~~~~~~~~~~~~~~~~~~~~~~~~~~~~~~~~~(10\mbox{\AA}\leq a\leq 50\mbox{\AA}) \eqno (18a)$$
$$Q_{iabs}^s(5)=a\Bigl(0.3\nu^2-0.00434\nu +0.000532\Bigr), ~~~(5\mbox{\AA}\leq a\leq 10\mbox{\AA}) \eqno(18b)$$
where the subscript $iabs$ used in $Q$ indicates absorption efficiency for
 ionic PAHs. In the far-ultraviolet and ultraviolet regions, the 
absorption and scattering efficiencies for neutral and ionic PAHs are 
identical. They, however, differ in visible and infrared regions as can be 
seen from Figure 3 which depicts this difference for $a=10$\AA~ and $a=50$\AA~. 
The curves have been plotted using above formulas. The curves show that the 
absorption by the ionic PAHs is generally greater in comparison to the neutral 
PAHs. More recently, Cecchi-Pestellini [16] have reported on variation of 
spectral properties of PAHs in various charged states. However, in the 
present work we limit ourselves only to the gross classification of neutral and 
charged PAHs, which is sufficient for our purposes. Scattering of 
electromagnetic radiation by a charged sphere has been treated by Klacka and 
Kocifaj [17] employing a generalization of the Mie theory.
\section{Extinction spectral features of PAHs} 
Having obtained the absorption and scattering efficiency formulas in the 
various frequency ranges, one can generate the extinction spectra for PAHs for 
a specified particle concentration $N_{PAH}$ and a size distribution 
function $f_p(a)$ over a radius range $[a_0,a_m]$:
5\AA~$\leq a_0,~ a_m\leq 50$\AA~ using the expression: 
$$K_{ext}(\lambda)=\frac{\pi N_{PAH}}{10^{10}}\int_{a_0}^{a_m} Q_{ext}^{PAH}(x)a^2 f_p(a)~ da, \eqno(19)$$
$f_p(a)$ being normalized to unity. In (19), $Q_{ext}^{PAH}(x)= 
Q_{abs}^{PAH}+Q_{sca}^{PAH}$ is the extinction 
efficiency of an individual scatterer of size parameter $x=2\pi a/\lambda$;  
$a$ and $\lambda$ being reckoned in unit $10^{-5}$ cm. This choice of units 
is in keeping with 
our earlier formulas for extinction spectra of silicate and graphite grains 
expressed in [1,2]. It may be observed that $Q_{sca}^{PAH}\ll Q_{abs}^{PAH}$ 
in the wavelength and radius ranges considered here. Hence, in (19) one can 
simply use $Q_{abs}^{PAH}$ in the place of $Q_{ext}^{PAH}$ without incurring 
appreciable error.\\\\
The $K_{ext}$ spectra (19) assumes much simpler form in the wavelength region 
where $Q_{abs}^{PAH}$ is linear in $a$. We observe that such is the case with 
the FUV and UV regions. In each of these regions we have 
$Q_{ext}^{PAH}\approx a\phi (\nu)$.  As a result, the corresponding expression 
for $K_{ext}^{PAH}$ in the 
wavelength region 1000\AA~$\leq\lambda\leq$5000\AA~ is given by the simple 
form 
$$K_{ext}^{PAH}(\lambda)\approx \frac{\pi N_{PAH}}{10^{10}}\overline{a^3}_{PAH}~ \phi(\nu), 
\eqno(20)$$
The volume corresponding to $N_{PAH}$ number of particles will be 
$$V_{PAH}=\frac{4\pi}{3}~ N_{PAH}~\overline{a^3}_{PAH}\times 10^{-15}, \eqno(21)$$
where 
$$\overline{a^3}_{PAH}=\int_{a_0}^{a_m}~ a^3~f_p(a)~ da, $$ 
$a$, $a_0$, $a_m$ all are in $10^{-5} cm. $ units. Using (21) in (20), we 
finally have, 
$$K_{ext}^{PAH}(\lambda)\approx \frac{3\times 10^5}{4}~V_{PAH}~\phi (\nu). \eqno (22)$$
In the other wavelength 
regions covering 5000\AA~$\leq\lambda\leq$22,500\AA~, however, the 
$K_{ext}^{PAH}$ spectra would be dependent on the size distributional details 
in more involved manner rather than simply being proportional to the third 
moment $\overline{a^3}$ as seen in (20). \\\\
If a population of PAHs specified by $N_{PAH}$, $f_p(a)$ and $V_{PAH}$ is 
included in the carbonaceous matter sector alongside graphite {parallel as 
well as perpendicular components} then due to stronger absorption features 
in the FUV and UV regions, a smaller amount of PAH (as compared to graphite) 
has the ability to increase the extinction contribution of the carbonaceous 
component of the dust by much larger amount. As a result, when one has to 
include a population of PAHs to any silicate-graphite model (two-component MRN 
model) in a 
manner so that the relative mass abundance criterion ( silicate mass/
carbonaceous mass) is maintained while the desired changes in the dust (PAHs 
+ silicate + graphite) extinction spectra are obtained, one can concentrate 
mainly to the FUV and UV regions. Thus among many, a simple option while 
extending a two-component silicate-graphite model would be to replace a definite 
amount of existing graphite (by mass) with equal amount (by mass)  of PAHs. 
Taking the mass density $(\approx 2.24~gm/cm^3)$ to be the same for both 
graphite and 
PAHs in bulk, we are thus led to a comparative study of the extinction spectra 
produced by equal volumes of graphite grains (parallel, perpendicular) 
and the PAHs. In the following, we examine this point in some details with 
respect to the MRN model considered by us in [1,2].\\\\

It may be recalled that in the two component graphite-silicate model of Mathis et al. (1979) [18], the  
spherical grains (graphite as well as silicate) are taken to follow a 
power-law size distribution
 of the form:  
$$f(a) \propto a^{-3.5}~~~~~~a_0\leq a \leq a_m,  \eqno(23)$$
 where, $a$ is the radius of the grain varying within the chosen values of   
$a_0$ and $ a_m$. Besides,  graphite material is taken to be present in two 
distinct structural varieties within the specified range of $a_0$ and $a_m$. 
This plausibility lies in the fact that graphite is a highly anisotropic 
material. The refractive index of graphite, therefore, depends on the 
orientation of electric field relative to its structural symmetry. Owing to 
practical difficulties in calculations of exact scattering (extinction) 
quantities due to the anisotropy, researchers have taken resort to
 an approximation known as $``\frac{1}{3}-\frac{2}{3}"$ approximation [19]. In 
this approximation, graphite grains are represented as a mixed population of 
isotropic 
spheres, of which $\approx\frac{1}{3}$ fraction have refractive index 
$m=m_{\parallel}$ 
(referred to as graphite parallel) and $\approx\frac{2}{3}$ fraction have the refractive 
index $m=m_{\perp}$ (referred to as graphite perpendicular). By virtue of 
[1,2], we have formulas for generating graphite extinction spectra for both 
the components ( graphite perpendicular and graphite parallel) within the 
parameter ranges 
$$0.002\mu m\leq a_0\leq 005\mu m;~~~0.15\mu m\leq a_m\leq 0.25\mu m.$$
By choosing $N_{gra}=4.4\times 10^8$ as the particle 
concentration for  both varieties of graphite (perpendicular as well as parallel) and 
the size range $[0.005\mu m:~ 0.25\mu m]$, the corresponding extinction spectra 
are displayed in Figure 4(b).\\

\begin{figure}[!h]
\centerline{\includegraphics[width=5.0in, height=10.0cm]{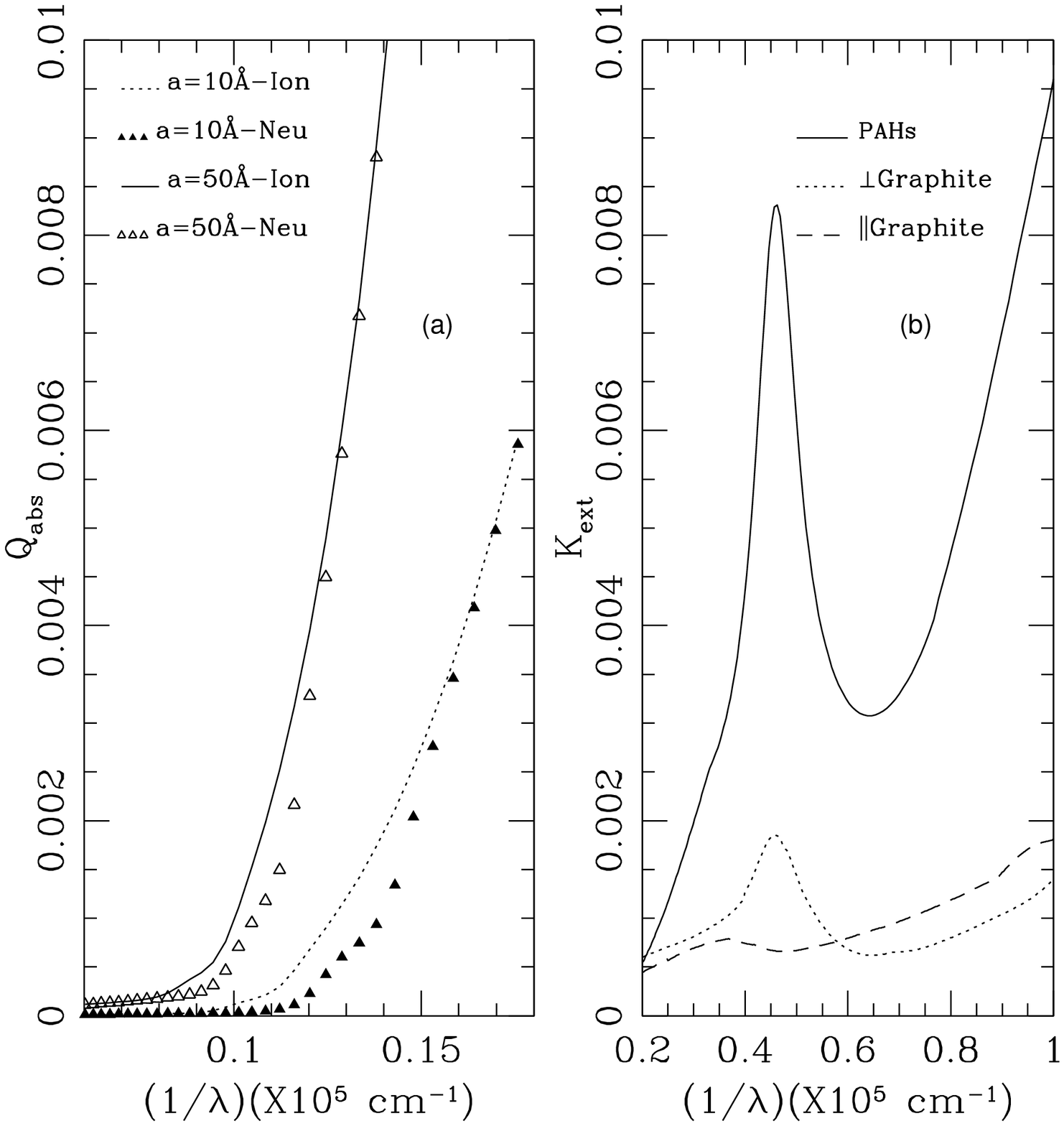}}
\begin{footnotesize}Fig 4 (a): Comparison of predictions of absorption efficiencies of Neutral and Ionic PAH's.\\
(b): Extincition spectra for equal volumes of PAH's and, parallel and perpendicular graphite grains.
 \end{footnotesize}
\end{figure}

The associated material volume of each of the 
graphite components is 
$V_{gra}=(4\pi/3)~ N_{gra}~ \overline{a^3}_{gra}\times 10^{-15} cc$, $\bar{a^3_{gra}}$ being the 
mean cube radius of the corresponding population of graphite particles. The 
radius $a$ has units in $10^{-5} cm;$ $\overline{a^3}_{gra}$ is to be evaluated using the formula 
[1] 
$$\overline{a^3}_{gra}=5a_0^3(n-1)\biggl(1-\frac{1}{n^5}\biggr)^{-1}, \eqno(24)$$
with $a_0=0.05$; $a_m=2.5$ and $n^2=(a_m/a_0)=50$.
\\\\
If we have to generate the PAHs extinction spectra corresponding to an equal 
bulk of graphite (perpendicular or parallel) having this fixed volume $V_{gra}$
we must use this value for $V_{PAH}$ occurring in (22). The resultant PAHs 
extinction spectrum is also displayed in Figure 4(b) along with that of the two 
graphite components bringing out the contrast in the individual extinction 
spectra produced by equal volumes of PAHs and the two varieties of graphite. It 
is clear from Figure 4(b) that the contribution of some mass of PAHs to the
 interstellar dust extinction will be  much greater in 
comparison to equal mass of graphite in the wavelength range  
1000\AA~$\leq \lambda\leq$ 5000\AA~ (FUV and UV regions). Further, if one 
looks at the ratio  $K_{ext}(\lambda)/K_{ext}(\lambda_v)$ fixing 
$\lambda_v$ at 5000\AA~  and compares among the three species viz. PAHs, 
graphite (perpendicular) and graphite (parallel), 
one can see easily that the ratio is much  greater for PAHs as compared to that 
of the other two species within the wavelength range 1000\AA~ to 5000\AA. 
Clearly, replacement of a small amount of graphite by an equal amount of PAHs 
in a dust mixture will cause an increase in the dust extinction throughout the 
wavelength range 1000\AA~ to 5000\AA. However, this increase would be  greater 
in the FUV region than that in the UV region. This is how the addition of PAHs 
to silicate-graphite model will help one to adjust the interstellar dust 
extinction spectral features in the FUV, UV regions.\\\\
In connection with our choice for the PAHs grain size range, we would like to 
mention that,  in Li and Draine [9] size range of PAHs has been taken to be 
$0.00035\leq a \leq 0.01~\mu m$ in their investigations. The absorption cross 
section of carbonaceous grains is represented as [9]
$$ C_{abs}^{carb}(a,\lambda)=\xi_{PAH}C_{abs}^{PAH}(a,\lambda) + 
(1-\xi_{PAH})C_{abs}^{gra}(a,\lambda),$$
$$\xi_{PAH}(a)=(1-q_{gra})\times min[1,(a_{\xi}/a)^3],$$
$$a_{\xi}=50\AA~,~~~~~~q_{gra}=0.01,$$
where $C_{abs}^{PAH},~ C_{abs}^{gra}$ are, respectively, absorption cross-sections
of PAH and graphite grains of radius $a$ and wavelength $\lambda$; 
$a_{\xi}$ is the grain radius from which the transition from PAH properties to 
graphite properties begins; $\xi_{PAH}$ is the PAH weight factor which falls off 
continually as $a$ increases from $a_{\xi}$ to infinity. However, as has been 
mentioned in [9], this choice of the weight factor is quite arbitrary. With 
this freedom of choice for  $\xi_{PAH}$, we consider a dust model in which the carbonaceous
 component is made up of PAHs of size range $5$\AA~$\leq a\leq 50$\AA~ and graphite grains in the size 
range $0.005\mu m\leq a\leq 0.25\mu m$.  
\section{Conclusions and discussions}
In this work, we have analyzed the extinction spectra generated by small size 
carbonaceous matter present in interstellar dust in the form of 
Polycyclic Aromatic Hydrocarbons (PAHs). Our investigations cover the 
wavelength region 1000\AA~ $\leq \lambda\leq$22,500\AA~ with PAHs size range 
5\AA~$\leq a\leq$50\AA~, $a$ being the radius of the PAH particle. In doing 
so, we have made use of the scattering and absorption efficiency data available
in [14] to obtain simple and accurate analytic formulas for scattering and 
absorption efficiencies expressible as functions of $a$ and $\lambda$. 
These formulas can be used expediently for computing extinction spectra 
generated by some bulk PAHs material specified through any selected size 
distribution. Thus, this work, together with our earlier work [1,2] essentially  completes our search for an analytic platform for interstellar dust extinction 
spectra analysis which employ MRN type dust models containing silicate, 
graphite and PAHs as well. \\\\
The scattering efficiency has $(a/\lambda)^4$ behavior over 
1800\AA~$\leq \lambda\leq$22,500\AA. For 1000\AA~$\leq\lambda\leq$1800\AA,  
the behavior is slightly different in that additional $(a/\lambda)^6$ and 
$(a/\lambda)^7$ behavior also creeps in. This is quite in keeping with the 
Pendorf formula [15] for the scattering efficiency in the case of small values 
of $a/\lambda$. The absorption efficiency has $a/\lambda$ behavior over 
1000\AA~$\leq\lambda\leq$5000\AA which is again in keeping with the Pendorf 
formula for absorption efficiency in respect of small values of $a/\lambda$. 
However, 
for $\lambda\geq$5000\AA~ sharper absorption features characterizing the 
special dispersive properties depending on size of the particle appear. This 
results in the absorption efficiency having a functional form which differs 
from that of Pendorf. A good account of modelling for these sharper absorption
 features occurring in the extinction spectra data experimentally generated by 
small sized hydrocarbons prepared in the laboratory is given by Li and Draine [9]. \\\\
The ionic PAHs show a slightly different extinction behavior as compared to 
their neutral counterparts. Both the neutral and the ionic species have the 
same scattering efficiencies and hence scattering properties of these are 
exactly the same within the size range 3.5\AA~$\leq a\leq$50\AA~ and the 
wavelength range 1000\AA~$\leq\lambda\leq$22,500\AA. The neutral and the ionic 
PAHs absorption efficiencies are same for 1000\AA~$\leq\lambda\leq$4000\AA; but 
they differ in the wavelength range 4000\AA~$\leq\lambda\leq$22,500\AA.
 The pattern of variation is illustrated in Figure 4(a). One can see that the 
range of wavelength [$\lambda_1,\lambda_2$] over which the absorption 
efficiency differs depends on the PAH radius $a$. It appears that both 
$\lambda_1$ and $\lambda_2$ increase with $a$. As the efficiency differences 
are not substantial, the preference of ionic over neutral PAHs to be included 
in the dust model does not seem to provide with much changes in the 
corresponding extinction spectra. \\\\
To estimate the extent to which the extinction spectra of a dust can be 
affected by having a definite amount of PAHs in it, we have compared the 
extinction spectra generated by equal volumes of all the three carbonaceous 
components viz, graphite (perpendicular), graphite (parallel) and PAHs. As 
PAHs extinction is essentially due to absorption only, we have chosen the 
wavelength range 1000\AA~$\leq\lambda\leq$5000\AA~ due to the simple form of the absorption 
efficiency. The comparison brings out very clearly the fact that PAHs indeed 
can act as a very effective carbonaceous component in MRN type dust models. It 
will be our future endeavor to use the analytic framework which includes this 
work alongside [1,2] for the analysis of extinction spectra data corresponding to MW, LMC and SMC.
  
\newpage
{\bf \Large Appendix}\\\\
The absorption cross section per $C$ atom has been denoted in [9] as 
$C_{abs}^{PAH}/N_C$, where $N_C$ is the number of $C$ atoms. The absorption 
efficiency, $Q_{abs}^{PAH}$, for a PAH of size $a$ can be expressed as  
$$Q_{abs}^{PAH}(a,\lambda)=\frac{N_C}{\pi a^2} C_{abs}^{PAH}/N_C$$
The ``radius" $a$ of a PAH containing $N_C$ number of $C$ atoms is defined to be the 
radius of a sphere with carbon density of graphite containing the same number 
of $C$ atoms, i.e., $a=1.286N_C^{1/3}$. Thus, 
$$Q_{abs}^{PAH}(a,\lambda)=\frac{a}{\pi (1.286)^3} C_{abs}^{PAH}/N_C ,$$
where $C_{abs}^{PAH}/N_C$ is given in equations (5)-(11) in [9] for various 
spectral ranges.
\clearpage
\noindent
{\Large\bf References}\\\\
$[1]$ Roy AK, Sharma SK and Gupta R, A study of frequency and size distribution
 dependence of extinction for astronomical silicate and graphite grains. J Quant
 Spectrosc Radiat Transf, 2009; 110, 1733-1740.\\  
$[2]$ Roy AK, Sharma SK and Gupta R, Frequency and size distribution  
dependence of visible and infrared extinction for astronomical silicate and 
graphite grains. J Quant Spectrosc Radiat Transf, 2010; 111, 795-801.\\  
$[3]$ Mathis JS, Rumpl W and Nordsieck KH, The size distribution of 
interstellar grains. ApJ, 1977; 217, 425-433.\\ 
$[4]$ Cardelli JA, Clayton GC and Mathis JS, The relationship between infrared,
 optical and ultraviolet extinction, ApJ, 1989; 345, 245-256.\\
$[5]$ Fitzpatrick EL and Massa D, An analysis of the shapes of interstellar 
extinction curves. V. The IR-through-UV curve morphology, ApJ, 2007; 663, 
320-341.\\ 
$[6]$ Platt, J. R. ApJ, 1956; 123, 486.\\
$[7]$ Donn, B. ApJ, 1968; 152, L129.\\
$[8]$ Draine, B. T. Li, A. 2007, ApJ, 657; 810.\\
$[9]$ Li, A. and Draine, B. T. Infrared emission from interstellar dust. II. The 
diffuse interstellar medium. ApJ, 2001; 554, 778-802.\\ 
$[10]$ Li, A. and Draine, B. T., ApJ, 2002a; 564, 803. \\
$[11]$ Weingartner, J.C. and Draine, B.T., ApJ, 2001a; 548, 296.\\
$[12]$ Li A., and Greenberg J. M., ' In Dust we Trust: an Overview of Observations and Theories of Interstellar Dust ',
 in Solid state Astrochemistry, 2003, Pirronello V,Krelowski J. \& Manico G. (Eds.), Kluwer, pp. 37-84.\\
$[13]$ Draine BT, Astronomical models of PAHS and dust. EAS Publ Ser 
2011; 46: 29-42.\\ 
$[14]$ Draine BT. At http://www.astro.princeton.edu/draine.\\
$[15]$ Pendorf RB, Scattering and extinction coefficients for small absorbing 
and nonabsorbing aerosols. J. Opt. Soc. Am., 1962; 52, 896-904.\\
$[16]$ Cecchi-Pestellini C, Malloci G, Mulas G, Joblin C and Wlliams DA, The 
role of the charge state of PAHs in ultraviolet extinction. A\&A, 3(2008); 486, L25\\
$[17]$ Klacka J and Kocijaj M, On the scattering of electromagnetic waves by a 
charged sphere. Prog. Electromagn Res., 2010; 17, 17-35.\\
$[18]$ Mathis J, The size distribution of interstellar particles. II - Polarization. ApJ, 1979; 232, 747-753.\\
{$[19]$ Draine BT and Malhotra S, On graphite and the 2175\AA~ extinction 
profile. ApJ. 1993; 414, 632-645.\\

\clearpage
\end{document}